\documentclass[twocolumn,showpacs,prl]{revtex4}
\usepackage{graphicx}
\newcommand{\ua}{\ensuremath{\uparrow}}
\newcommand{\da}{\ensuremath{\downarrow}}

\newcommand{\ttt}{$2 \times 2\times 2$}
\begin{document}

\title{Electronic Structure and Origin of Ferromagnetism in
CaB$_6$}

\author{Zachary M. Helms}
\affiliation{
Department of Physics and Center for High Performance Simulation,
North Carolina State University, Raleigh, NC 27695.}

\author{Prasenjit Sen}
\affiliation{Harish-Chandra Research Institute, Chhatnag Road, Jhunsi,
Allahabad 211019, INDIA.}

\author{Lubos Mitas}
\affiliation{
Department of Physics and Center for High Performance Simulation,
North Carolina State University, Raleigh, NC.}

\date{\today}

\begin{abstract}
Electronic structure calculations using quantum Monte Carlo (QMC)
methods conclusively show that pure CaB$_6$ is a narrow-gap 
semiconductor
with an $X$-point gap $\sim 1.3$ eV. This should put to rest
controversies as to whether the compound is a semimetal or a
semiconductor, as also theories suggesting it to be an excitonic
insulator. Our extensive Hartree-Fock (HF) and density functional
theory (DFT) calculations support the view
  that the hexaboride
ferromagnetism
can be induced by contamination with magnetic element atoms like Fe and
that
presence of La is not essential for this. La
impurity, however, gives rise to a metallic state with a
small electron-like Fermi surface as seen in ARPES experiments.
\end{abstract}

\maketitle

Recently, a great deal of interest has been generated in divalent
hexaboride CaB$_6$
mainly due to observation of unusual ferromagnetism (FM) in the
system supposedly with Lanthanum doping~\cite{LaFM}. 
The unusual feature of this ferromagnetic
state is that it has very small moment ($\sim 0.07$ $\mu_B$ per La
atom) and yet has very high Curie temperature, T$_{\rm c}\sim 600$K.
Theoretical models with unconventional many-body mechanisms
were proposed as explanation for this FM in
absence of $d$- or $f$- electrons in the system. For example,
it was suggested \cite{LaFM, Ceperley} that doped hexaboride forms 
the magnetically ordered ground state of a dilute electron gas
while Zhitomirsky {\it et al.} and other workers~\cite{excins}
suggested it to be an excitonic insulator. 
Central to the idea of excitonic insulator is the assumption of a
small
band overlap of top of the B $p$ valence states and bottom of the Ca 
$d$
conduction states. This band overlap was suggested by early 
density functional theory (DFT) band
structure calculations~\cite{LDAband}. It is well-known by now that
DFT in local density approximation 
seriously underestimates the band gap, 
hence models based on a small 
band
overlap in these calculations are somewhat
questionable. Going beyond DFT, the perturbational GW method
band structures, which incorporate
self-energy corrections, have been calculated for
the system. Though one of them predicts a large 
band gap at the X-point of
the Brillouin zone (BZ) of this simple cubic system~\cite{GWgap}, 
another shows a small overlap~\cite{GWoverlap}.
A very recent calculation using weighted density approximation 
predicts a
band gap of $\sim 0.8$ eV at the X-point~\cite{WDA}.
These conflicting results clearly point out that our understanding
of basic properties
of this interesting material is still rather limited.

Theoretical scenario being unclear as it is, there is considerable
controversy regarding metallic/semiconducting character of CaB$_6$
in the experimental results, too. Semi-metallic nature is supported by
observation of metallic resistivity~\cite{metallicR1,metallicR2}, 
the de Haas-van Alphen
effect~\cite{metallicR1,dHvA}, and Shubnikov-de Haas 
results~\cite{SdH}. 
On the other hand, NMR~\cite{NMR},
angle resolved photoemission spectroscopy (ARPES)~\cite{ARPES}, and
thermopower experiments~\cite{thermop} suggest that CaB$_6$ is a 
narrow-gap semiconductor. Some recent high resolution 
ARPES~\cite{highARPES} and combined transport, optical
and tunneling measurements~\cite{combo} very strongly suggest 
that pure CaB$_6$ is a
semiconductor, while metallic character and FM in the system originate
from boron-related defects.

While the pure system is poorly understood, there is very little
understanding of the origin of FM in the system. 
Authors in ref.~\cite{Monnier} 
proposed a scenario for FM in La-doped samples
over a small concentration range by studying 
formation energies and associated moments of various defects.
However, given the large formation energy
of B$_6$ octahedron vacancy they found, and observation of FM in 
systems
even without La-doping, make their proposal less likely. Various
experimental groups have suggested, and there is growing consensus on
this point, that FM in CaB$_6$ is an extrinsic effect related to
magnetic impurities such as Fe and Ni~\cite{extrns1,extrns2,Bennet}, 
probably coming from B used in
sample preparation. The most convincing argument for this is
provided by the experiments of Cho {\it et. al}~\cite{combo}.

Hoping to clarify some of these issues, we did a detailed first 
principles electronic structure calculation of pure CaB$_6$
using Hartree-Fock (HF), DFT and quantum Monte Carlo  (QMC)
methods, and also did 
an analysis of CaB$_6$ with defects with DFT and HF approaches.
Defects include substitutional impurities--Fe and La in
the cation sublattice, and vacancies--a single missing B.
For the first time, through the use of
explicitly correlated many-body wavefunctions in variational Monte 
Carlo
(VMC) and diffusion Monte Carlo (DMC) calculations,
we conclusively show that pure CaB$_6$ is a semiconductor with an
$X$-point gap of $\sim 1.3$ eV. Using HF and 
DFT methods, we have shown that i) La
is not essential for ferromagnetism in the system, while ii) Fe
substitutional impurities give rise to ferromagnetism, iii) 
B vacancy, and Fe and La substitutional impurities make the system 
metallic, iv) CaB$_6$ with La substitutional impurities has a small
electron-like Fermi surface (FS),
explaining some of the observations in ARPES
experiments~\cite{ARPES,highARPES}.
Since experimental information about atomic structure around
these defects is essentially absent, our studies 
should be understood as an exploration of various possible scenarios.
Nevertheless, we believe our results suggest 
explanations to most of the experimental 
observations at least qualitatively, and
give an accurate quantitative estimate of the gap in the pure system, 
thus
clarifying the overall physics of CaB$_6$.

We performed our HF and DFT calculations using
CRYSTAL98 and CRYSTAL03 codes~\cite{crystal98,crystal2003}. 
We used both all electron basis
sets and ECPs in our calculations and they gave similar results. For 
QMC
calculations we used the same basis sets with ECPs as used in HF.
For QMC calculations, we used a Slater-Jastrow wavefunction of the
form
\begin{equation}
\Psi = D^{\ua } D^{\da }exp\left [\sum_{i,j,I}u(r_{ij},r_{iI}) \right 
]. 
\end{equation}
The Jastrow correlation part depends on
electron-ion  distances $r_{iI}$ and electron-electron distances 
$r_{ij}$
and the term includes electron-electron
cusp terms and altogether contains 10 variational parameters. 
In the first
step, these variational parameters are optimized by minimizing the
variance in the local energy, $E_L = \frac{H\Psi}{\Psi}$, $H$ being
the Hamiltonian of the system in which the core electrons are replaced 
by pseudopotentials (for Ca the Ne-core electrons were eliminated). 
Finally, a large part of the variational bias is eliminated by 
fixed-node DMC. Details of these methods have been discussed in the 
literature~\cite{RMP}.

\begin{figure}%[h]
\scalebox{0.3}{ \includegraphics{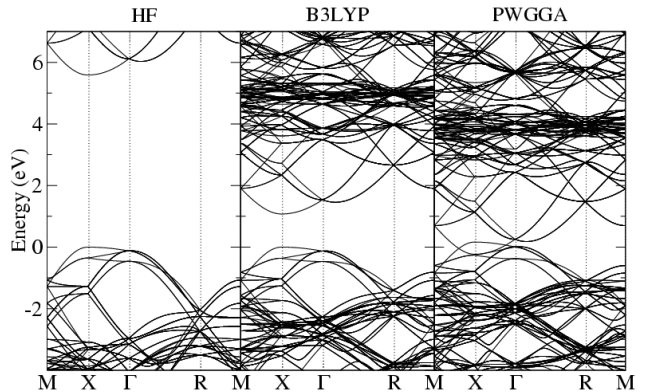} }
\hfill
\caption{ Band structure of pure CaB$_6$ calculated with HF, 
B3LYP functional and PW91 GGA functional.}
\label{fig:pureplot}
\end{figure}

{\em Pure CaB$_6$}:
As is typical with these methods, HF gave a large X-point gap of $\sim 
5.50$
eV, while the use of hybrid functional B3LYP reduced the gap to $\sim 
0.86$
eV. PW91 generalized gradient approximation (GGA)
gave a conducting state and a small
band overlap at the X-point, as found in all other such calculations in 
the literature (see fig.~\ref{fig:pureplot}). 
For QMC calculations we used a $2\times 2\times 2$
supercell so that the X-point of the primitive BZ is folded on to
the $\Gamma$-point of the BZ corresponding to the simulation cell.
Since we are interested in the gap at the $X$-point of the primitive 
BZ, we
can work with wavefunctions at the $\Gamma$-point of the supercell
BZ. 
The one-particle orbitals for the Slater part of the many-body 
wavefunction for our QMC calculations were taken from HF 
calculations. For the ground state wavefunction, the $N$ lowest energy
orbitals were included. For the excited state, we promoted one
electron from the top of the valence band to the bottom
of the conduction band at the $\Gamma$-point of simulation BZ which
   corresponds to the folded X-point.
Variational principle guarantees that the calculated energy is a
strict
upper bound for the lowest energy state of the symmetry imposed by the 
trial
function. 
   Since we are interested in direct gap, the excited state of
interest in not the ground state of the particular symmetry (which is
determined by the $k$-point occupation in a solid). It is well known 
that in such
cases the DMC energy is not necessarily 
a strict upper bound of the excited state
energy. However, fixed-node DMC gives a good estimate of the excited 
state
energy. In fact, it is exact if the nodal structure of the trial 
function
$\Psi(R)$ is exact, and in most practical cases the fixed-node
constraint is dominant and leads to energies which are higher than the 
exact energies
for a given state (see
ref.~\cite{andy}, for example).
We optimized the variational parameters
in the Jastrow part only for the ground state wavefunction, and used 
the
same set for the excited state. The precise value of the gap we 
obtained
in our DMC calculations is 1.3(2) eV. This is in very good agreement
with all the experiments that give an estimate of the bulk
band-gap~\cite{ARPES,highARPES,combo}. Obviously, this result is based
on fixed-node DMC single determinant trial wavefunctions.

Our results based on the fixed-node single-determinant trial wave 
function
DMC method, which is a many-body approach different from
previous calculations, suggest 
that the pure CaB$_6$ 
compound is a narrow-gap semiconductor. Hence we conjecture
that the metallic character
seen in transport measurements 
have originated from defects in
the system. This result also suggests that the 
theories with unconventional mechanisms explaining
FM in this system using early DFT band structures
might not be correct and we believe that our calculations support the
arguments for extrinsic origins. Therefore, we follow with
  calculations of CaB$_6$ with defects,
proposing explanations for the origin of FM and metallic behavior as
observed. As we mentioned, experimental information 
about atomic structures of possible 
defects in hexaborides is not available. Therefore our 
calculations of the hexaboride with defects only serve as models to 
illustrate 
the qualitative impact of imperfections on the electronic structure.

{\em CaB$_6$ with defects}:
We present here results from our calculations of three different 
defects in
CaB$_6$. We consider substitution of Ca in the lattice by Fe and La, 
and an
isolated B vacancy. The first one is motivated by the suggestions
that magnetic ions, possibly Fe or Ni, are responsible for FM in this
system. Although there have been indications that Fe and Ni reside 
close to
the surface layer~\cite{Bennet}, there are also experiments suggesting 
that they penetrate substantially into the bulk~\cite{Meegoda}. We 
treat these
as bulk
substitutional impurities to avoid complications of surface effects
such as relaxation and reconstruction in our calculations.
It has also been speculated that FM with high T$_c$ is due to 
boride formation--FeB or Fe$_2$B--which have T$_c$'s of 598 K and 1015 
K
respectively. However, this has been contradicted by other
works~\cite{extrns1}. Since the experimental evidence is not 
conclusive, 
we do not consider any possible boride formation either.
B vacancy would presumably be the most natural defect in a
nominally B deficient sample.
Although no FM was seen in B deficient samples in ref.~\cite{Monnier},
we still study this system to understand its electronic properties.
La substitution is motivated by the original observation of FM in
the La substituted compound Ca$_{1-x}$La$_x$B$_6$.

Concentrations of various defects in conditions of thermal equilibrium
depend on their formation energies (FE). The defect formation energy 
is defined in the usual way,
as the difference between the binding energies of the pure solid and 
the 
solid with defect. Since there are no hints about structural 
relaxations
around defects from experiments we did not consider 
such relaxations. At least in some
cases these have been shown to be
negligible~\cite{Monnier}.  However, such relaxations cannot be ruled
out {\it a priori} and therefore our
  formation energies (Table~\ref{table:formE}) are most probably biased
towards the defect formation being less favorable.
Nevertheless, since we are interested in a basic, qualitative 
understanding
of the electronic structure changes in presence of defects, we believe 
that our model calculations are sufficient for this purpose.

\begin{table}
\caption{Defect formation energies in eV calculated with HF and B3LYP.}
\begin{tabular}{lccccccccccc}
Defect           & & & & & & & HF    & & & B3LYP \\  \hline \hline
Fe-substitution  & & & & & & & 15.6 & & & 6.9 \\
B-vacancy        & & & & & & & 15.8 & & & 14.8 \\
La-substitution  & & & & & & & -0.5 & & & -2.8 \\  \hline
\end{tabular}
\label{table:formE}
\end{table}

Any Fe substitutional impurity in the system, coming from 
impurities in boron, or by some
other means during sample preparation, would be present in minute 
amounts
because of its relatively large FE. The smallest impurity
concentration we have been able to study is $\sim 3.7 \%$ of Fe doping
(one Ca replaced in a $3\times 3\times 3$~ supercell) because of
system size limitations. Going down from 12.5 \% to
3.7 \%, an Fe impurity always stabilizes a FM state over a
paramagnetic one. As an origin of this FM state, we note that there
is a splitting of Fe $d$ states due to the ligand field of the B 
octahedra.
In presence of this cubic field, the five degenerate $d$ states split into
two different energy levels: triply degenerate t$_{2g}$ states, and doubly
degenerate e$_g$ states. The result is a $\frac{3}{2}$ spin on Fe with a
$-\frac{1}{2}$ spin over surrounding B atoms that forms a local magnetic
moment. A FM state will form when these local moments interact either through
hybridized Fe-B states, or in regions where there is a high Fe
concentration, such as near the surface.

\begin{figure}%[h]
\scalebox{0.30}{ \includegraphics{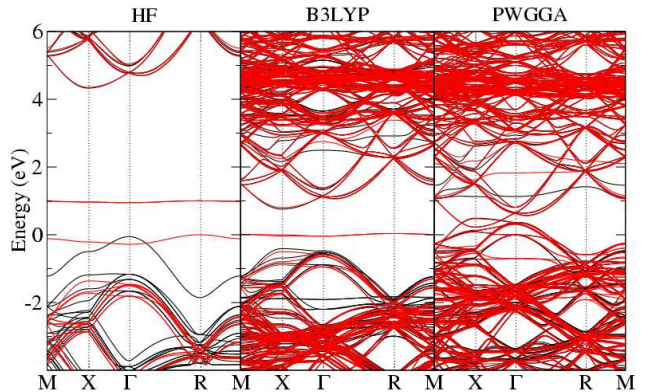} }
\hfill
\caption{(Color online) Band structure of CaB$_6$ with 3.7\% Fe 
impurity 
with HF, B3LYP and PW91
functionals. Black(solid) and red(dotted) lines represent up and down 
spin 
energies in the FM ground state. Fermi energy in all cases has been set
to $0$.}
\label{fig:feplot}
\end{figure}

Fig.~\ref{fig:feplot} shows the band structure 
plots for a CaB$_6$ system with a Fe impurity, calculated
with HF, standard B3LYP and PW91 functionals. 
From the band structure, the system is seen to be metallic at the
B3LYP level. From the position of the Fermi level 
it is also clear that the system would have a small hole-like FS. 
We consider these results qualitative in nature since the concentration
of the iron defects is artificially high due to the size of the
supercell. Nevertheless, the presence of the Fe localized moment will
be similar in smaller concentrations since the $t_{2g}$ states are 
quite deep and show very little hybridization with the neighboring 
boron states. 
The most important changes one can expect might be actually in the
minority spin-channel where larger supercells might enable better
state relaxations with possibly larger spin-density compensation
resulting in overall smaller spin polarization.

Next we consider a B vacancy in a \ttt~ system which amounts to 2.08\%
impurity concentration. Removing a B has the effect of taking away a B 
$p$ valence electron from top of the valence band. 
The system becomes metallic and the position of the Fermi energy tells 
that it
has a hole-like FS at the $\Gamma$-point of the BZ of the \ttt~ cell.
Although our calculations show that a magnetically ordered state is
favored, the energy difference with a paramagnetic state is only $\sim 
0.05$ eV
per primitive cell. This energy difference is too small to be taken
as an indication of a magnetic ground state, and in fact,
can be an artifact of the B3LYP method. This conjecture is supported
by the
GGA results in ref.~\cite{Monnier} where B vacancy is not seen to give 
rise to any appreciable magnetic moment.

\vspace*{.1in}
\begin{figure}
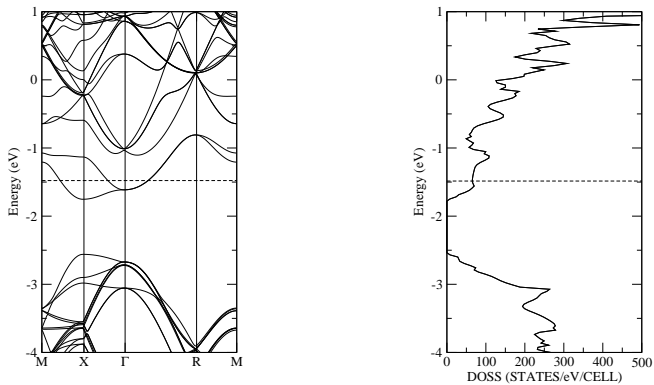
%[t]
\scalebox{0.30}{ \includegraphics{fig3a} }
\hfill
\scalebox{0.30}{ \includegraphics{fig3b} }
\caption{B3LYP band structure and density of states of 3.7\% La-doped
CaB$_6$. The Fermi energy is indicated by the dashed line.}
\label{fig:laplot}
\end{figure}

Now we come to our study of La substitution of Ca. 
Because of the initial observation of magnetism in 
Ca$_{1-x}$La$_x$B$_6$,
it was thought that La plays a role in the origin of FM in the system. 
This
makes the study of La substitution particularly interesting. Also note
that the formation energy
  of a La substitutional impurity turns out to be negative
indicating that any La atoms present during sample preparation would
readily get incorporated into CaB$_6$. In our HF
calculations we find that La-doped CaB$_6$, down to an impurity
concentration of 1.6\%, has a paramagnetic ground state.
This is found to be the case also at the B3LYP level of approximation. 
In any case, since HF is known to favor a high-spin state, one can
reasonably expect the La-doped system to be nonmagnetic. 
This, along with our prediction of FM in Fe doped CaB$_6$, lends
support to the claims that FM in this system is an extrinsic effect due 
to
some magnetic atoms, and is not related to the presence of La.
In Fig.~\ref{fig:laplot} we show the
band structure and density of states for a system with 3.7\%
La-doping.  The system is found to be metallic. From the position of 
the Fermi 
surface we find that the system has a small electron-like FS at the
$\Gamma$-point in the BZ. Comparing band structure in 
Fig.~\ref{fig:laplot}
with that of a $(3\times 3\times 3)$ supercell (not shown) of pure
CaB$_6$, we find that there is an upward shift of the Fermi energy
with only minor changes in energy bands. Thus La replacing a Ca atom
acts as an electron donor in the system. We believe that the
electron-like Fermi surface seen in ARPES 
measurements~\cite{ARPES,highARPES} 
might have originated from La substitutional impurities.

Finally we study the effects of a B vacancy and a Fe substitutional 
impurity 
being present together in CaB$_6$.
This is motivated by the observation in ref.~\cite{combo} that
although
CaB$_6$, CaB$_{6+\delta}$ and Ca$_{1-x}$La$_x$B$_6$ prepared with
lower
purity B exhibit FM, no such effect is seen in CaB$_{6-\delta}$. This
suggests that although Fe substitutional impurities give rise to FM, B
vacancies neutralize that effect and when both are present in the 
system it
is no longer FM. This is precisely borne out by our calculations.
We have studied two extreme scenarios as far as the positions of the B
vacancy and Fe impurity are concerned--i) they are on neighboring
sites, and ii) they are as far apart as possible in our $3 \times 3
\times 3$ simulation cell. In both cases a nonmagnetic state is
preferred over a FM state in our HF calculations. Again, HF typically
preferring a high-spin state, we can take this as a reasonable
indication that a B vacancy destroys the FM state created by a Fe
impurity.

In conclusion, using explicitly correlated many-body wavefunctions, we
have conclusively shown that pure CaB$_6$ is a narrow-gap semiconductor
with an $X$-point gap of $1.3(2)$ eV. Our HF and DFT calculations show 
that possible imperfections in the
crystal--B vacancy, Fe or La substitutional impurities--drive the
system metallic, explaining the apparent anomalies in the experimental
results. Out of these, La substitutional impurities give rise to a 
small
electron-like Fermi surface, which was seen in experiments but had no
clear explanation. We have also shown that FM in this system is related
to magnetic impurities, {\it e.g.}, Fe and probably has no relation 
with
the presence of La. Another interesting fact emerging from our
calculations, explaining the fact that FM was never seen in 
CaB$_{6-\delta}$ samples in Ref.~\cite{combo}, is
that a B vacancy destroys any magnetic moment created by Fe impurities.

We gratefully acknowledge the support by ONR-N00014-01-1-0408,
and NSF DMR-0121361, DMR-0121361 grants and
the computer time allocations at NCSA and PSC  facilities, and the Kabir
cluster at HRI.

\end{document}